\documentclass[12 pt]{article}
\usepackage{amssymb,amsmath,amsfonts,amssymb,bm,stmaryrd}
\usepackage{graphicx,color,colortbl,texdraw,enumerate}
\usepackage{enumerate}
\usepackage[parfill]{parskip}

\usepackage{amssymb, color}

\input{tcilatex}
\begin{document}

\title{Is There a Bubble in LinkedIn's Stock Price?}
\author{Robert Jarrow\thanks{%
Johnson Graduate School of Management, Cornell University, Ithaca, NY, 14853
and Kamakura Corporation.} \and Younes Kchia\thanks{%
Centre de Math\'{e}matiques Appliqu\'{e}es, Ecole Polytechnique, Paris.}
\and Philip Protter\thanks{%
Statistics Department, Columbia University, New York, NY, 10027.} \thanks{
Supported in part by NSF grant DMS-0906995}}
\date{\today }
\maketitle

\begin{abstract}
Recent academic work has developed a method to determine, \emph{in real time}%
, if a given stock is exhibiting a price bubble. Currently there is
speculation in the financial press concerning the existence of a price
bubble in the aftermath of the recent IPO of LinkedIn. We analyze stock
price tick data from the short lifetime of this stock through May 24, 2011,
and we find that LinkedIn has a price bubble.
\end{abstract}

\section{Introduction and Summary}

It has been well documented in the financial press that a methodology is
needed that can identify an asset price bubble in real time. William Dudley,
the President of the New York Federal Reserve, in an interview with Planet
Money~\cite{planet} stated \textquotedblleft ...what I am proposing is that
we try to identify bubbles in real time, try to develop tools to address
those bubbles, try to use those tools when appropriate to limit the size of
those bubbles and, therefore, try to limit the damage when those bubbles
burst.\textquotedblright 

It is also widely recognized that this is not an easy task. Indeed, in 2009
the Federal Reserve Chairman Ben Bernanke said in Congressional Testimony~%
\cite{Bernanke} \textquotedblleft It is extraordinarily difficult in real
time to know if an asset price is appropriate or not\textquotedblright .
Without a quantitative procedure, experts often have different opinions
about the existence of price bubbles. A famous example is the oil price
bubble of 2007/2008. Nobel prize winning economist Paul Krugman wrote in the
New York Times that it was not a bubble, and two days later Ben Stein wrote
in the same paper that it was.

Although not yet widely known by the finance industry, the authors have
recently developed a procedure based on a sophisticated mathematical model
for detecting asset price bubbles \emph{in real time} (see~\cite{JKP}). We
have successfully back-tested our methodology showing the existence of price
bubbles in various stocks during the dot-com era of 2000 to 2002. We also showed that some stocks in that period that might have suspected of being bubbles, were in fact not. But, we
have not yet tested our method on stocks in real time. That is, we have not
tested them until now.

Inspired by a New York Times article~\cite{NYT} discussing whether or not in
the aftermath of the LinkedIn IPO the stock price had a bubble, we obtained
stock price tick data from Bloomberg. And, we used our methodology to test
whether LinkedIn's stock price is exhibiting a bubble. \textit{We have
found, definitively, that there is a price bubble!}

Our method consists of assuming a general (and generally accepted) evolution
for the stock price, estimating its volatility using state of the art
estimators, and then extrapolating the volatility function to see if a
certain calculus integral based on the volatility is finite or infinite. If
it is finite, the stock price has a bubble; if the integral in question is infinite then the stock is not undergoing bubble pricing; our
test is indeterminate in a small set of cases where the volatility tends to
infinity at a rate between both possibilities. In the case of LinkedIn, the
volatility function is well inside the bubble region. There is no doubt
about its existence.

Our results can be summarized in the following graph showing an
extrapolation of an estimated volatility function for LinkedIn's stock
price. Put simply, the theory developed in~\cite{JPS1},\cite{JPS2}, and~\cite%
{JKP} tells us that if the graph of the volatility versus the stock price
tends to infinity at a faster rate than does the graph of $f(x)=x$, then we
have a bubble. Below we have a graph of the volatility coefficient of
LinkedIn together with its extrapolation, and the reader can clearly see
that the graph indicates the stock has a price bubble.

\begin{figure}[h]
\begin{center}
\includegraphics[scale=0.57]{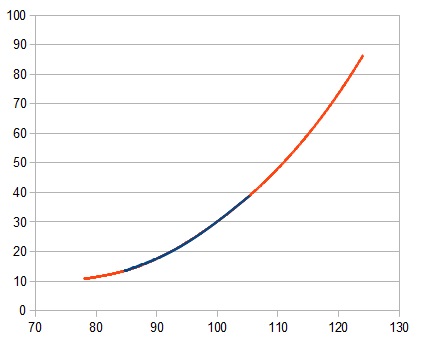}
\end{center}
\caption{Estimation and Extrapolation of the Volatility Function}
\label{RKHS2}
\end{figure}
The blue part of the graph is the estimated function, and the red part is
its extrapolation using the technique of Reproducing Kernel Hilbert Spaces
(RKHS).

These LinkedIn results illustrate the usefulness of our new methodology for
detecting bubbles in real time. Our methodology provides a solution to the
problems stated by both Chairman Bernanke and President Dudley, and it is
our hope that they will prove useful to regulators, policy makers, and
industry participants alike.

\bigskip

\noindent \textbf{Acknowledgement:} The authors thank Peter Carr and Arun
Verma for help in obtaining quickly the tick data for the stock LinkedIn.

\section*{The Empirical Test}

In this section we first review the methodology contained in \cite{JPS2} to
determine whether or not LinkedIn's stock price is experiencing a bubble,
and then we apply this methodology to LinkedIn minute by minute stock price
tick data obtained from Bloomberg. We conclude that LinedIn's stock is
indeed experiencing a price bubble.

\subsection{The Methodology}

The methodology is based on studying the characteristics of LinkedIn's stock
price process. If LinkedIn's stock price is experiencing a bubble, it can be
shown \cite{JPS2} that the stock price's volatility will be unusually large.
Our empirical methodology validates whether or not LinkedIn's stock
volatility is large enough.

To perform this validation, we start by assuming that the stock price
follows a stochastic differential equation of a form that makes it a
diffusion in an incomplete market (see (\ref{1}) below). The assumed stock
price evolution is very general and fits most stock price processes
reasonably well. For identifying a price bubble, the key characteristic of
this evolution is that the stock price's volatility $\sigma (S_{t})$ is a
function of the level of the stock price.

Next, we use sophisticated methods to estimate the volatility coefficient $%
\sigma (x)$. Since the data is necessarily finite, we can estimate the
values of $\sigma (x)$ only on the set of stock prices observed (which
is a compact subset of $\mathbb{R}_{+}$) We use two methods to extend $%
\sigma $ to all of $\mathbb{R}_{+}$. One, we use parametric methods combined
with a comparison theorem. Two, we use an indexed family of Reproducing
Kernel Hilbert Spaces (RKHS) combined with an optimization over the index
set to obtain the best possible extension given the data (this is a sort of
bootstrap procedure).

This \textquotedblleft knowledge of $\sigma (S_{t})$\textquotedblright\ then
enables us to determine whether the stock price is a martingale or is a
strict local martingale under any of an infinite collection of risk neutral
measures. If it is a strict local martingale for all of the risk neutral
measures (which corresponds to a certain calculus integral being finite by
Feller's test for explosions), then we can conclude that the stock price is
undergoing a bubble. Otherwise, there is not a stock price bubble.

\subsection{The Estimation}

We assume that LinkedIn's stock price
is a diffusion of the form 
\begin{eqnarray}
dS_{t}&=&\sigma (S_{t})dW_{t} +b(S_t,Y_t)dt\\ \label{1}
dY_t&=&s(Y_t)dB_t+g(Y_t)dt
\end{eqnarray}%
where $W$ and $B$ are independent Brownian motions. This model permits that under the physical probability measure, the
stock price can have a drift that depends on additional randomness, making
the market incomplete (see \cite{JPS2}). Nevertheless for this family of models, $S$ satisfies the following equation for every neutral measure:
$$
dS_{t}=\sigma (S_{t})dW_{t}
$$
Under this evolution, the stock price exhibits a bubble if and only if 
\begin{equation}
\int_{a}^{\infty }\frac{x}{\sigma ^{2}(x)}dx<\infty \ \ for\ all\ \ a>0.
\end{equation}%
We test to see if this integral is finite or not.

\begin{figure}[h]
\begin{center}
\includegraphics[scale=0.62]{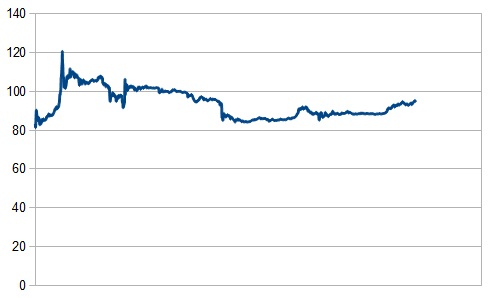}
\end{center}
\caption{LinkedIn Stock Prices from 5/19/2011 to 5/24/2011.(The observation
interval is one minute.)}
\label{LinkedInSpot}
\end{figure}

To perform this test, we obtained minute by minute stock price tick data for
the 4 business days 5/19/2011 to 5/24/2011 from Bloomberg. There are exactly
1535 price observations in this data set. The time series plot of LinkedIn's
stock price is contained in Figure \ref{LinkedInSpot}. The prices used are
the open prices of each minute but the results are not sensitive to using
open, high or lowest minute prices instead.

The maximum stock price attained by LinkedIn during this period is \$120.74
and the minimum price was \$81.24. As evidenced in this diagram, LinkedIn
experienced a dramatic price rise in its early trading. This suggests an
unusually large stock price volatility over this short time period and
perhaps a price bubble.

\begin{figure}[h]
\begin{center}
\includegraphics[scale=0.65]{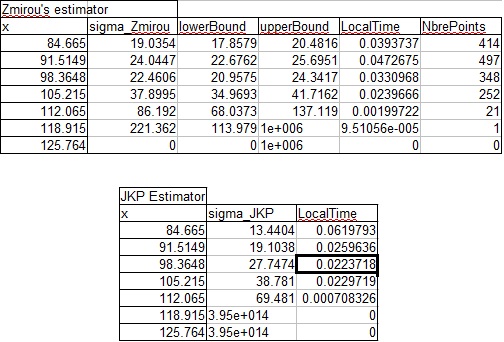}
\end{center}
\caption{Non-parametric Volatility Estimates.}
\label{Estimation}
\end{figure}

Our bubble testing methodology first requires us to estimate the volatility
function $\sigma $ using local time based non-parametric estimators. We use
two such estimators. We compare the estimation results obtained using both
Zmirou's estimator (see Theorem 1 in \cite{JKP}) and the estimator developed
in \cite{JKP} (see Theorem 3 in the same reference). The implementation of
these estimators requires a grid step $h_{n}$ tending to zero, such that $%
nh_{n}\rightarrow \infty $ and $nh_{n}^{4}\rightarrow 0$ for the former
estimator, and $nh_{n}^{2}\rightarrow \infty $ for the later one. We choose
the step size $h_{n}=\frac{1}{n^{\frac{1}{3}}}$ so that all of these
conditions are simultaneously satisfied. This implies a grid of 7 points.
The statistics are displayed in figure \ref{Estimation}.

Since the neighborhoods of the grid points \$118.915 and \$125.764 are
either not visited or visited only once, we do not have reliable estimates
at these points. Therefore, we restrict ourselves to the grid containing
only the first five points. We note that the last point in the new grid
\$112.065 still has only been visited very few times.

\begin{figure}[h]
\begin{center}
\includegraphics[scale=0.65]{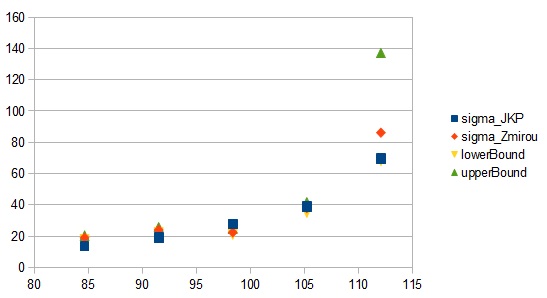}
\end{center}
\caption{Non parametric Volatility Estimation Results.}
\label{EstimationGraph}
\end{figure}

When using Zmirou's estimator, confidence intervals are provided. The
confidence intervals are quite wide. Given these observations, we apply our
methodology twice. In the first test, we use a 5 point grid. In the second
test, we remove the fifth point where the estimation is uncertain and we use
a 4 point grid instead. The graph in figure \ref{EstimationGraph} plots the
estimated volatilities for the grid points together with the confidence
intervals.

\begin{figure}[h]
\begin{center}
\includegraphics[scale=0.65]{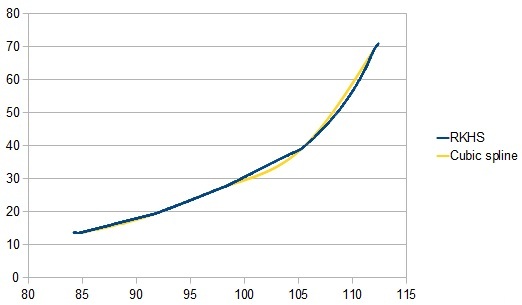}
\end{center}
\caption{Interpolated Volatility using Cubic Splines and the RKHS Theory}
\label{Interpolation}
\end{figure}

The next step in our procedure is to interpolate the shape of the volatility
function between these grid points. We use the estimations from our non
parametric estimator with the 5 point grid case. For the volatility time
scale, we let the 4 day time interval correspond to one unit of time. This
scaling does not affect the conclusions of this paper. When interpolating
one can use any reasonable method. We use both cubic splines and reproducing
kernel Hilbert spaces as suggested in \cite{JKP}, subsection 5.2.3 item
(ii). The interpolated functions are in figure \ref{Interpolation}.

From these,we select the kernel function $K_{1,\tau }$ as defined in Lemma
10 in \cite{JKP}, and we choose the parameter $\tau =6$.

The next step is to extrapolate the interpolated function $\sigma ^{b}$ using
the RKHS theory to the left and right stock price tails. Define $f(x)=\frac{1%
}{\sigma ^{2}(x)}$ and define the Hilbert space 
\begin{equation*}
H_{n}=H_{n}\big(\lbrack 0,\infty \lbrack \big)=\big\{f\in C^{n}\big(\lbrack
0,\infty \lbrack \big)\mid \lim_{x\rightarrow \infty }x^{k}f^{(k)}(x)=0\text{
for all }0\leq k\leq n-1\big\}
\end{equation*}%
where $n$ is the assumed degree of smoothness of $f$. We also need to define
an inner product. A smooth reproducing kernel $q^{RP}(x,y)$ can be
constructed explicitly (see Proposition 2 in \cite{JKP}) via the choice 
\begin{equation*}
\langle f,g\rangle _{n,m}=\int_{0}^{\infty }\frac{y^{n}f^{(n)}(y)}{n!}\frac{%
y^{n}g^{(n)}(y)}{n!}\frac{dy}{w(y)}
\end{equation*}%
where $w(y)=\frac{1}{y^{m}}$ is an asymptotic weighting function. We
consider the family of RKHS $H_{n,m}=(H_{n},\langle ,\rangle _{n,m})$, in
which case the explicit form of $q_{n,m}^{RP}$ is provided in Proposition 2
in \cite{JKP} in terms of the Beta and the Gauss's hypergeometric functions.

\begin{figure}[h]
\begin{center}
\includegraphics[scale=0.65]{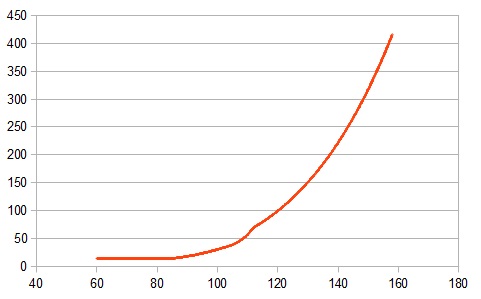}
\end{center}
\caption{RKHS Based Extrapolation of $\protect\sigma ^{b}$}
\label{RKHS}
\end{figure}

For $n\in \{1,2\}$ fixed, we construct our extrapolation $\sigma =\sigma _{m}
$ as in \cite{JKP}, 5.2.3 item (iv), by choosing the asymptotic weighting
function parameter $m$ such that $f_{m}=\frac{1}{\sigma _{m}^{2}}$ is in $%
H_{n,m}$, $\sigma _{m}$ exactly matches the points obtained from the non
parametric estimation, and $\sigma _{m}$ is as close (in norm 2) to $\sigma
^{b}$ on the last third of the bounded interval where $\sigma ^{b}$ is
defined. Because of the observed kink and the obvious change in the rate of
increase of $\sigma ^{b}$ at the forth point, we choose $n=1$ in our
numerical procedure. The result is shown in figure \ref{RKHS}.

We obtain $m=9.42$.

From Proposition 3 in \cite{JKP}, the asymptotic behavior of $\sigma $ is
given by 
\begin{equation*}
\lim_{x\rightarrow \infty }x^{m+1}f(x)=n^{2}B(m+1,n)\sum_{i=1}^{M}c_{i}
\end{equation*}%
where $M=5$ is the number of observations available, $B$ is the Beta
function, and the coefficients $(c_{i})_{1\leq i\leq M}$ are obtained by
solving the system 
\begin{equation*}
\sum_{i=1}^{M}c_{i}q_{n,m}^{RP}(x_{i},x_{k})=f(x_{k})\text{ for all }1\leq
k\leq M
\end{equation*}%
where $(x_{i})_{1\leq i\leq M}$ is the grid of the non parametric
estimation, $f(x_{k})=\frac{1}{\sigma ^{2}(x_{k})}$ and $\sigma (x_{k})$ is
the value at the grid point $x_{k}$ obtained from the non-parametric
estimation procedure. This implies that $\sigma $ is asymptotically
equivalent to a function proportional to $x^{\alpha }$ with $\alpha =\frac{%
1+m}{2}$, that is $\alpha =5.21$. This value appears very large, however, the
proportionality constant is also large. The $c_{i}$'s are automatically
adjusted to exactly match the input points $(x_{i},f(x_{i}))_{1\leq i\leq M}$%
.

We plot below the functions with different asymptotic weighting parameters $m
$ obtained using the RKHS extrapolation method, without optimization. All
the functions exactly match the non-parametrically estimated points.

\begin{figure}[h]
\begin{center}
\includegraphics[scale=0.65]{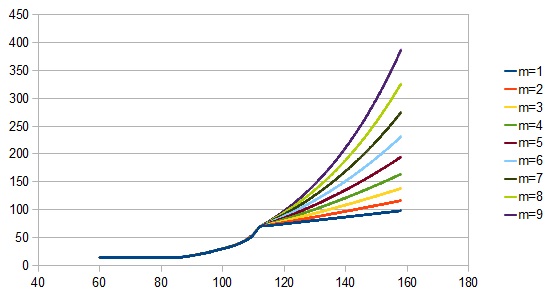}
\end{center}
\caption{Extrapolated Volatility Functions using Different Reproducing
Kernels}
\label{RKHSAll}
\end{figure}

The asymptotic weighting function's parameter $m=9.42$ obtained by
optimization appears in figure \ref{RKHSAll} to be the estimate most
consistent (within all the functions, in any Hilbert Space of the form $%
H_{1,m}$, that exactly match the input data) with a "natural" extension of
the behavior of $\sigma ^{b}$ to $\mathbb{R}^{+}$. \textit{The power }$%
\alpha =5.21$\textit{\ implies then that LinkedIn stock price is currently
exhibiting a bubble.}

Since there is a large standard error for the volatility estimate at the end
point \$112.065, we remove this point from the grid and repeat our
procedure. Also, the rate of increase of the function between the last two
last points appears large, and we do not want the volatility's behavior to
follow solely from this fact. Hence, we check to see if we can conclude
there is a price bubble based only on the first 4 reliable observation
points. We plot in figure \ref{RKHS2} the function $\sigma ^{b}$ (in blue)
and its extrapolation to $\mathbb{R}^{+}$, $\sigma $ (in red).

\begin{figure}[h]
\begin{center}
\includegraphics[scale=0.65]{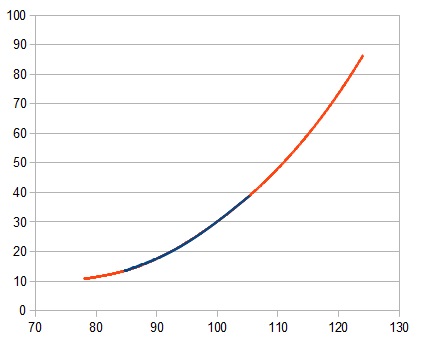}
\end{center}
\caption{RKHS based Extrapolation of $\protect\sigma ^{b}$}
\label{RKHS2}
\end{figure}

Now $M=4$. With this new grid, we can assume a higher regularity $n=2$ and
we obtain, after optimization, $m=7.8543$. This leads to the power $\alpha
=4.42715$ for the asymptotic behavior of the volatility. Again, although
this power appears to be high given the numerical values $%
(x_{k},f(x_{k}))_{1\leq k\leq 4}$, the coefficients $(c_{i})_{1\leq i\leq 4}$
and hence the constant of proportionality are adjusted to exactly match the
input points. The extrapolated function obtained is the most consistent
(within all the functions, in any $H_{2,m}$, that exactly match the input
data) in terms of extending 'naturally' the behavior of $\sigma ^{b}$ to $%
\mathbb{R}^{+}$. Again, we can conclude that there is a stock price bubble.

\end{document}